\documentclass[11pt]{article}

\usepackage[preprint]{acl}

\usepackage{times}
\usepackage{latexsym}

\usepackage[T1]{fontenc}

\usepackage[utf8]{inputenc}

\usepackage{microtype}

\usepackage{inconsolata}

\usepackage{graphicx}

\usepackage{xspace}
\usepackage{algorithm}
\usepackage{algpseudocode}
\usepackage{amsmath}
\usepackage{multirow}
\usepackage{booktabs}
\usepackage{verbatim}
\usepackage[table]{xcolor}
\usepackage{siunitx}
\definecolor{rowgray}{gray}{0.94}
\definecolor{llmgray}{gray}{0.88}

\title{SVR-MAD: A Bayesian-Inspired Framework for Posterior-Guided Multi-Agent Debate}

\author{
  \textbf{Weifan Jiang},
  \textbf{Rana Shahout},
  \textbf{Minghao Li},
  \textbf{Zhenting Qi},
  \textbf{Yilun Du},
\\
  \textbf{Michael Mitzenmacher},
  \textbf{Minlan Yu}
\\
  Harvard University
\\
  \url{https://github.com/weifanjiang/SVR-MAD}
}

\newcommand{\secref}[1]{\S\ref{#1}}
\newcommand{\sysname}{SVR-MAD\xspace}

\newenvironment{packeditemize}{\begin{list}{$\bullet$}{\setlength{\itemsep}{1pt}\addtolength{\labelwidth}{-4pt}\setlength{\leftmargin}{\labelwidth}\setlength{\listparindent}{\parindent}\setlength{\parsep}{1pt}\setlength{\topsep}{2pt}}}{\end{list}}

\begin{document}
\maketitle

\begin{abstract}
Multi-Agent Debate (MAD) improves LLM-agent accuracy but suffers from rapid context growth, limiting scalability in larger multi-agent settings. Existing methods prune low-utility communications using prior signals, such as token-level log-likelihoods or LLM self-reported confidence. However, these signals become unreliable under hallucination, degrading the accuracy of MAD methods that rely on them. We propose \sysname, a Bayesian-inspired MAD framework that treats pre-debate signals as priors and debate outcomes as posterior-style evidence for estimating agent correctness. \sysname uses this evidence to incrementally construct the communication graph, prioritizing agents whose answers survive peer challenges. Experiments across multiple LLMs and benchmarks show that \sysname reduces token cost by up to 61\% while matching or improving accuracy relative to the most accurate competing MAD baseline.
\end{abstract}

\section{Introduction}

Multi-Agent Debate (MAD) improves the accuracy of Large Language Model (LLM) agents across diverse domains, including research, mathematics, and coding~\citep{Du2024Improving, Li2023Camel, Wu2024Autogen}. By exchanging intermediate reasoning steps and answers across rounds, agents can revise their responses based on peer information and achieve accuracy gains beyond single-agent settings. However, MAD becomes costly as the communication graph grows. Each shared message is appended to another agent’s context, increasing the input-token cost and KV cache usage. With more agents or debate rounds, this overhead compounds quickly, limiting scalability.

Recent work reduces the cost of MAD by removing nodes (i.e., participating agents) and/or edges (i.e., communication links between agents) from all-to-all communication graphs~\citep{Liu2025Groupdebate, Chen2025Sid, Zeng2025S2Mad, Zhu2026Demystifying}. These methods typically leverage information available prior to the debate (e.g., token-level log-likelihoods, LLM's self-stated confidence) to identify and remove potentially low-utility communications. For example, outputs with high min. log-likelihood values are often considered to be reliable, and therefore are likely to be correct without further debate~\citep{Chen2025Sid}.

In this work, we uncover that although pre-debate signals are effective in common cases, their informativeness can collapse on challenging problem instances relative to the LLM’s capabilities. On such instances, incorrect agents may hallucinate with high confidence, while correct agents may remain uncertain~\citep{Kakai2025Why}. Methods based on prior signals can prematurely exclude agents or communication links that would otherwise be beneficial, thereby reducing MAD's accuracy gains.

To address this limitation, we take a Bayesian-inspired view of MAD: prior agent signals can be unreliable under hallucination, whereas debate outcomes provide posterior evidence of reliability. We focus on whether an agent preserves its answer after incorporating peer messages. Intuitively, an agent supported by sound reasoning is less likely to be convinced by flawed peer arguments. We propose \textit{Survival Rate} (SVR), which measures how often an agent retains its original belief after being challenged. Because SVR is grounded in observed post-debate behavior, it provides a more hallucination-robust signal for identifying trustworthy reasoning processes and answers.

We propose \sysname, an efficient MAD framework that incrementally builds a communication graph using posterior debate outcomes. \sysname initializes agent correctness scores from prior signals, probes $S$ communication links to high-prior receivers, and updates scores based on whether agents retain or revise their answers. It then prioritizes high-score agents in later communications and terminates once an agent’s score exceeds a threshold. This greedy strategy identifies reliable solutions early, reducing token costs. Across two LLMs and datasets, \sysname reduces token costs by up to 61\% while matching or improving accuracy relative to the most accurate competing MAD baseline.

\section{Background and Related Works}

\noindent\textbf{MAD preliminary.} In MAD, LLM agents iteratively exchange progress. Suppose $N$ participating agents. Let $C_{t}^i$ be the output (i.e., reasoning and answer) from agent $i$ at turn $t$. Then, in all-to-all MAD, agent $i$ generates the next-turn output, $C_{t+1}^i$, based on the following context:

[Question] + [turn $1,\dots,t-1$ history] + $C_t^i$ + $DebateFormat(C_{t}^1,\dots,C_t^{i-1}, C_t^{i+1}\dots C_t^N)$

where $DebateFormat$ is a string manipulation function that concatenates peer outputs and adds a prompting hint to guide agent $i$ to reflect on peer contexts, such as \textit{"These are the solutions from other agents [Insert peer solutions here]...Using the solutions from other agents as additional information, provide your answer..."}~\citep{Du2024Improving}.

While MAD improves LLM-agent accuracy, it scales poorly with more agents due to rapid context growth. Under all-to-all communication, each agent’s context grows as $\mathcal{O}(N^2t)$, increasing computation and KV-cache pressure while also risking long-context degradation from diluted salient information~\citep{Liu2024Lost, Laban2026Llm}.

\noindent\textbf{Efficient MAD solutions.} Existing works leverage pre-debate signals to improve MAD. One line of work uses agent-level reliability signals to identify agents whose answers already appear reliable and can therefore conclude without any debate~\citep{Chen2025Sid, Eo2025Debate}. Another line uses cross-agent similarity signals to remove communication links that appear redundant~\citep{Zeng2025S2Mad, Liu2025Groupdebate, Zhu2026Demystifying}.

\section{Analysis of Prior and Posterior Signals}

\noindent\textbf{Prior signals.} Prior signals are derived from agents’ initial responses. They typically measure output quality and are used to exclude agents whose initial responses already appear reliable from debate, and/or prune the communication graph to prioritize deliberation with high-quality agents, aiming to improve accuracy with less communication.

We study three prior signals: min. log-likelihood (min-LL), which captures the least likely token; perplexity (PPL), which measures response-level generation uncertainty; and the LLM's self-reported confidence (conf).

\noindent\textbf{Our proposal: SVR.} SVR measures how often an agent retains its initial belief after debating with peers. This evidence-grounded signal provides a posterior estimate of agent reliability that is less dependent on potentially hallucinated prior information. Let $D$, $r$, and $c$ denote the total number of debates involving an agent, the number of times the agent retains or changes its belief after debate, respectively. We define $SVR=(r-c) / D$.

\begin{figure}
\centering
\includegraphics[width=\columnwidth]{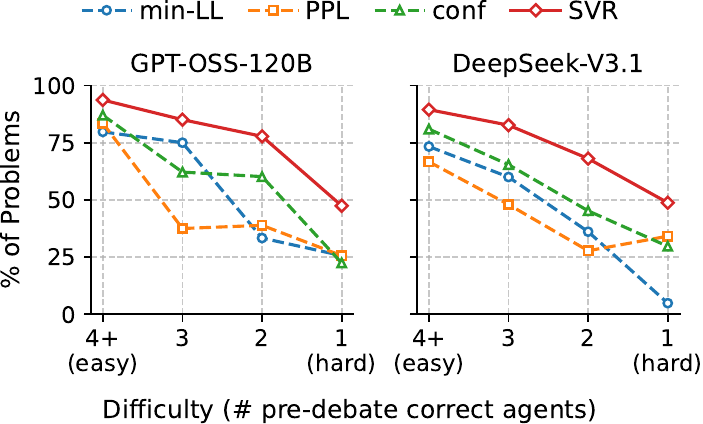}
\caption{\label{fig:motivation}\textbf{SVR better identifies correct agents than prior signals under increasing difficulty.} Percentage of problems where the top-ranked agent under each signal is correct, as a function of increasing difficulty.}
\end{figure}

\noindent\textbf{Approach.} We validate SVR on two LLMs: GPT-OSS-120B~\citep{Openai2025gptoss} and DeepSeek-V3.1~\citep{Deepseek2024Deepseek}, using questions from IMO-AnswerBench~\citep{Luong2025Towards}. For each question, we run six agents and group questions by difficulty, measured by the number of agents that are correct before debate. We compare three prior signals against SVR by measuring how often the top-ranked agent under each signal is correct.

\noindent\textbf{Results.} Figure~\ref{fig:motivation} reports this percentage across difficulty groups, showing how each signal aligns with agent correctness as problems become harder. From 4+ correct agents (easiest) to 1 correct agent (hardest), the correctness of top-ranked agents under min-LL, PPL, and confidence drops from 67--87\% to 5--34\%. \textbf{This sharp decline shows that prior signals become unreliable on difficult problems, where hallucinated or overconfident reasoning makes model behavior poorly aligned with correctness.} In contrast, SVR remains predictive as difficulty increases: among questions with only 2 and 1 correct agents, SVR achieves 68--78\% and 47--49\%, respectively, outperforming the strongest prior signal by 18--23 and 15--22 points.


\section{\sysname design}

\begin{algorithm}
\caption{\label{alg:sysname}\sysname}
\begin{algorithmic}[1]
\Statex \textbf{Input:} (1) Agents $\mathcal{A}_{1}\dots\mathcal{A}_{N}$, (2) Pre-debate answers $y_1\dots y_N$, (3) Communications per round $S$, (4) Budget $B_{max}$, (5) Acceptance threshold $\tau$
\Statex \textbf{Output:} Final answer $\hat{y}$

\State \textbf{Initialize} correctness scores

    $Corr[\mathcal{A}_i]=\text{Prior}(\mathcal{A}_i)$ for $1 \leq i \leq N$

\State \textbf{Initialize} $B=B_{max}$

\While{$B > 0$}
    \State $\mathcal{A}_{r} = \arg\max_{\mathcal{A}} \; Corr[\mathcal{A}]$
    \State $\text{Disagree} = \{\mathcal{A}_j : y_j \neq y_{r}\}$
    \State $\mathcal{P} = \text{top-}S \text{ of } \text{Disagree} \text{ ranked by } Corr$
    \ForAll{$\mathcal{A}_s \in \mathcal{P}$}
        \State $Outcome = Debate(\mathcal{A}_s \rightarrow \mathcal{A}_{r})$
        \State $Corr[\mathcal{A}_{r}] \mathrel{+}= \ell(Outcome)$
    \EndFor
    \If{$Corr[\mathcal{A}_{r}] \geq \tau$}
        \State \Return $\hat{y} = y_{r}$
    \EndIf
    \State $B \mathrel{-}= S$
\EndWhile

\State \Return Fallback majority-voting
\end{algorithmic}
\end{algorithm}

Algorithm~\ref{alg:sysname} presents \sysname's workflow. The inputs are the pre-debate states of all agents, including prior signals, reasoning tokens, and initial answers. \sysname has the following steps:

\noindent\textbf{Prior-based correctness score initialization} (line 1): We initialize each agent's correctness score using pre-debate prior signals, which provide an initial estimate of whether $A_i$ holds a correct answer. These scores are updated as debate outcomes become available.

\noindent\textbf{Incremental graph construction} (line 3-15): \sysname operates iteratively by selecting one agent ($A_{r}$) as the receiver at a time. It then pairs the receiver with $S$ peers and performs $S$ pairwise debates. The outcomes (i.e., whether $A_{r}$ retained or revised its belief after debate) are used to update $A_{r}$'s correctness score.

\noindent\textbf{Sender and receiver selection} (line 4-6): In each iteration, \sysname selects the agent with the highest correctness score as the receiver. This greedy strategy prioritizes agents most likely to hold correct answers, helping \sysname confirm reliable solutions and terminate with fewer communications. Among peers that disagree with the receiver $\mathcal{A}_{r}$, \sysname selects up to $S$ strongest challengers based on their correctness scores.

\noindent\textbf{Posterior-style score update} (line 7-9): \sysname updates each receiver's correctness score after every debate. For receiver $\mathcal{A}_{r}$, let $D$, $r$, and $c$ denote the running counts of observed debates, retentions, and changes, respectively. After each debate, we increment $D \mathrel{+}= 1$ and either $r \mathrel{+}= 1$ or $c \mathrel{+}= 1$ based on debate outcome. We then apply a posterior-dominant rule:

$Corr[A_r] =
\begin{cases}
SVR(D,r,c), & \text{if } D > 0,\\
Prior(A_r), & \text{otherwise.}
\end{cases}
$

Thus, prior signals guide selection only before debate evidence is available; once observed, SVR dominates and is refreshed after each debate.

\noindent\textbf{Early termination} (line 11-13): After each debate, \sysname checks whether the receiver's updated correctness score exceeds the acceptance threshold. If so, \sysname terminates the debate early and returns that agent's answer as the final solution to reduce overall communication cost.

\noindent\textbf{Fallback majority-voting} (line 16): If no agent commits within budget, each agent contributes one vote: the majority of its observed post-debate answers, with its pre-debate answer $y_i$ as tiebreaker. \sysname returns the majority over the $N$ votes, tie-breaking by the pre-debate majority among $y_1, \ldots, y_N$.

\section{\label{sec:evaluation}Evaluation}

\begin{table*}[h]
\centering
\small
\setlength{\tabcolsep}{5pt}
\renewcommand{\arraystretch}{1.12}
\begin{tabular}{
l l
S[table-format=2.1]
S[table-format=2.1]
S[table-format=2.1]
S[table-format=2.1]
S[table-format=2.1]
S[table-format=2.1]
}
\toprule
\textbf{LLM} & \textbf{Method}
& \multicolumn{3}{c}{\textbf{IMO-AnswerBench}}
& \multicolumn{3}{c}{\textbf{HLE}} \\
\cmidrule(lr){3-5}
\cmidrule(lr){6-8}
&
& {\textbf{NComm} $\downarrow$}
& {\textbf{Tok} $\mathbf{(\times 10^3)}$ $\downarrow$}
& {\textbf{Acc} $\uparrow$}
& {\textbf{NComm} $\downarrow$}
& {\textbf{Tok} $\mathbf{(\times 10^3)}$ $\downarrow$}
& {\textbf{Acc} $\uparrow$} \\
\midrule

\multirow{5}{*}{GPT-OSS-120B}
& \cellcolor{rowgray}Self Consistency
& \cellcolor{rowgray}0.0 & \cellcolor{rowgray}45.4 & \cellcolor{rowgray}36.7
& \cellcolor{rowgray}0.0 & \cellcolor{rowgray}14.1 & \cellcolor{rowgray}13.9 \\
& \cellcolor{white}GroupDebate
& \cellcolor{white}19.4 & \cellcolor{white}164.1 & \cellcolor{white}41.8
& \cellcolor{white}18.4 & \cellcolor{white}111.0 & \cellcolor{white}20.2 \\
& \cellcolor{rowgray}SID-ET
& \cellcolor{rowgray}17.5 & \cellcolor{rowgray}85.5 & \cellcolor{rowgray}38.7
& \cellcolor{rowgray}12.8 & \cellcolor{rowgray}39.6 & \cellcolor{rowgray}17.2 \\
& \cellcolor{white}S$^2$-MAD
& \cellcolor{white}22.8 & \cellcolor{white}131.4 & \cellcolor{white}42.1
& \cellcolor{white}14.0 & \cellcolor{white}67.7 & \cellcolor{white}21.0 \\
& \cellcolor{rowgray}\sysname
& \cellcolor{rowgray}{\bfseries 5.6} & \cellcolor{rowgray}{\bfseries 82.1} & \cellcolor{rowgray}{\bfseries 42.8}
& \cellcolor{rowgray}{\bfseries 7.3} & \cellcolor{rowgray}{\bfseries 39.1} & \cellcolor{rowgray}{\bfseries 21.4} \\

\midrule

\multirow{5}{*}{DeepSeek-V3.1}
& \cellcolor{white}Self Consistency
& \cellcolor{white}0.0 & \cellcolor{white}20.2 & \cellcolor{white}31.8
& \cellcolor{white}0.0 & \cellcolor{white}9.8 & \cellcolor{white}14.1 \\
& \cellcolor{rowgray}GroupDebate
& \cellcolor{rowgray}19.8 & \cellcolor{rowgray}234.6 & \cellcolor{rowgray}40.1
& \cellcolor{rowgray}19.6 & \cellcolor{rowgray}108.7 & \cellcolor{rowgray}14.9 \\
& \cellcolor{white}SID-ET
& \cellcolor{white}20.0 & \cellcolor{white}103.0 & \cellcolor{white}33.8
& \cellcolor{white}16.6 & \cellcolor{white}33.7 & \cellcolor{white}14.1 \\
& \cellcolor{rowgray}S$^2$-MAD
& \cellcolor{rowgray}20.9 & \cellcolor{rowgray}154.7 & \cellcolor{rowgray}36.8
& \cellcolor{rowgray}18.7 & \cellcolor{rowgray}71.7 & \cellcolor{rowgray}{\bfseries 16.7} \\
& \cellcolor{white}\sysname
& \cellcolor{white}{\bfseries 8.9} & \cellcolor{white}{\bfseries 91.7} & \cellcolor{white}{\bfseries 41.5}
& \cellcolor{white}{\bfseries 5.3} & \cellcolor{white}{\bfseries 30.3} & \cellcolor{white}{\bfseries 16.7} \\

\bottomrule
\end{tabular}
\vspace{3pt}
\caption{\textbf{\sysname achieves the best cost-accuracy tradeoff across all evaluated settings.} Evaluation results across two datasets and two LLMs.
Arrows indicate whether higher ($\uparrow$) or lower ($\downarrow$) metric values are better. Best results are bold (NComm/Tok comparisons exclude Self Consistency, which is a no-debate reference).}
\label{tab:evaluation}
\vspace{-8pt}
\end{table*}

\noindent\textbf{Baselines.} We compare \sysname to (1) Self Consistency~\citep{Wang2023Self}, which aggregates independent agent solutions by majority vote without debate; (2) GroupDebate~\citep{Liu2025Groupdebate}, which restricts full-context exchange to random subgroups and allows only final-answer messages across groups; (3) SID-ET~\citep{Chen2025Sid}, which uses token log-likelihood-based self-confidence to select agents for debate; and (4) S$^2$-MAD~\citep{Zeng2025S2Mad}, which prunes communication links between agents with similar answers or response embeddings. All methods use the same pre-debate responses per question.

\noindent\textbf{Metrics.} We capture the tradeoff between communication cost and accuracy using three metrics: (1) number of communications (NComm), which counts pairwise context transfers in the abstract communication graph;
(2) total tokens, which captures the concrete overhead under specific model, dataset, and prompt settings; and (3) accuracy.

\noindent\textbf{MAD setting.} We use 6 agents per problem. All methods run for at most two debate rounds and may terminate after round one if at least five agents reach a clear consensus, thereby avoiding unnecessary cost inflation.

\noindent\textbf{LLMs and datasets.} We evaluate on GPT-OSS-120B and DeepSeek-V3.1 using IMO-AnswerBench and a filtered HLE~\citep{Phan2026Benchmark} subset containing only text-only, non-math, multiple-choice questions. This filtering ensures compatibility with GPT-OSS-120B (text-only LLM), avoids domain overlap with IMO-AnswerBench, and enables automatic correctness evaluation. We skip questions where all six agents share a unanimous answer before debate.

\noindent\textbf{Additional experimental settings.} We include other experimental details in Appendix~\ref{sec:additional-exp-setting}.

\subsection{Main results}

\noindent\textbf{\sysname achieves the best cost-accuracy tradeoff across all evaluated settings.} \sysname matches or exceeds every MAD baseline on accuracy while reducing per-question communications by 48–75\% and total tokens by 38–61\%, relative to the most accurate MAD baseline. On three of the four settings, \sysname strictly improves accuracy, with gains of up to 1.8, 7.7, and 4.7 points compared to GroupDebate, SID, and S$^2$-MAD, respectively. On DeepSeek-V3.1/HLE, \sysname matches S$^2$-MAD's accuracy but uses 72\%/58\% fewer communications/tokens per question.

\begin{figure}
\centering
\includegraphics[width=0.9\columnwidth]{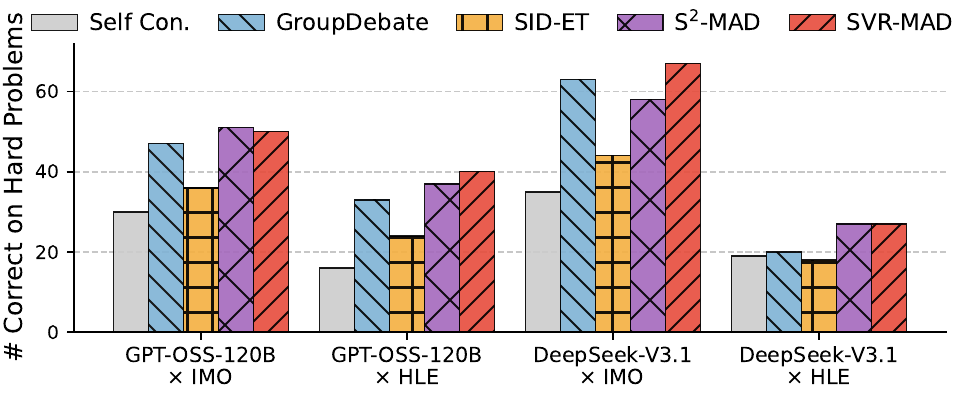}
\caption{\label{fig:tail-acc} \textbf{\sysname achieves high accuracy on hard problems across all evaluated settings.} Accuracy of each method on problems where the number of correct agents before debate is within 1-3.}
\end{figure}

\noindent\textbf{\sysname achieves robust accuracy on hard problems while remaining cost-efficient.} Figure~\ref{fig:tail-acc} shows the number of hard problems each method answers correctly, where hard problems have 1--3 correct agents before debate. \sysname matches or exceeds all baselines on three of four settings, trailing only by one question to S$^2$-MAD on GPT-OSS/IMO. As S$^2$-MAD incurs 1.4--2.5$\times$ higher token cost on these hard problems, \sysname remains the most robust and cost-efficient.

\noindent\textbf{Ablation studies.} Appendix~\ref{sec:ablation} isolates the contribution of SVR by replacing it with alternative posterior signals in Algorithm~\ref{alg:sysname}. SVR improves accuracy by up to 8.75 points at matched token cost compared to the best alternative, showing that SVR provides an effective posterior signal beyond the benefits of the communication control flow.

\section{Conclusion}

We present \sysname, a Bayesian-inspired MAD framework that leverages posterior evidence from debate outcomes. \sysname first probes selected communications to estimate agent reliability, then incrementally constructs the communication graph toward agents with stronger posterior signals. Evaluation across multiple LLMs and datasets shows \sysname improves the cost-accuracy trade-off, reducing communication links and tokens by up to 75\% and 61\%, while matching or improving accuracy relative to the most accurate competing MAD baseline.

\newpage
\section*{Limitations}

We discuss several limitations of this work. First, SVR is an evidence-grounded proxy for correctness. Our central assumption is that agents with sound, well-supported reasoning are more likely to retain their answers when challenged by peers. While our results show that SVR is more robust than prior signals on hard problems, it does not guarantee correctness. Incorrect agents may also retain their answers when peer arguments are weak or when the model is overconfident in an incorrect reasoning path. A higher acceptance threshold can partially mitigate this issue, and we plan to further improve SVR reliability in future work.

Second, our evaluation focuses on closed-ended reasoning tasks where final answers can be compared automatically. This setting enables consistent majority voting and correctness measurement, but it does not fully capture open-ended generation tasks, where determining answer equivalence is harder. We plan to extend \sysname to open-ended tasks in future work.

Third, the effectiveness of \sysname may depend on the debate prompt, the number of agents, and the model family. Different LLMs may respond differently to peer challenges: some may revise too aggressively, while others may retain answers even when presented with strong counterarguments. We plan to further improve \sysname's robustness by studying model-specific debate prompts and validating the method across more model families and domains.

\section*{Ethical considerations}

This work does not raise any ethical concerns.



\bibliography{custom}

\appendix

\section{\label{sec:additional-exp-setting}Additional Experimental Settings}

We provide additional experimental details to ensure full reproducibility of our work.

\noindent\textbf{Dataset pre-processing.} As described in \secref{sec:evaluation}, we filter HLE to text-only, non-math, multiple-choice questions. For each <LLM, dataset> setting, we remove questions on which all six agents give the same answer before debate, since debate is unnecessary in such cases. Note that we do keep questions where all agents are wrong but disagree on the answer.

\begin{table}[h]
\centering
\begin{tabular}{lcc}
\toprule
 & IMO & HLE \\
\midrule
GPT-OSS-120B & 297 & 267 \\
DeepSeek-v3.1             & 299 & 276 \\
\bottomrule
\end{tabular}
\caption{\label{tab:dataset-size}Valid sample size per experimental setting after all pre-processing steps.}
\end{table}

The questions removed from each <LLM, dataset> setting differ, so Table~\ref{tab:evaluation} is not suitable for cross-setting comparisons of the same MAD method. However, the same set of questions is used for all MAD methods under each <LLM, dataset>. Table~\ref{tab:dataset-size} records the valid sample size per evaluation setting.

\noindent\textbf{LLM settings.} We keep default sampling parameters for all settings: temperature 1.0, top-p 0.95, top-k 40. We set the maximum output tokens to 16,384 per LLM response. We set the reasoning effort to medium for both LLMs, as higher efforts often produce long reasoning chains that exhaust the maximum token cap before outputting a valid answer.

\noindent\textbf{GroupDebate.} We consider two possible settings: two groups of three (3-3), and three groups of two (2-2-2). The original paper suggests that (3-3) yields higher accuracy while (2-2-2) uses fewer tokens (Table 2 in~\citep{Liu2025Groupdebate}). Therefore, we tested (3-3) first, and planned to switch to (2-2-2) if (3-3) yields better accuracy than \sysname. The GroupDebate results in Table~\ref{tab:evaluation} are all using (3-3). We did not further evaluate (2-2-2), as (3-3) already trails \sysname in accuracy among all <LLM, dataset> settings.

\noindent\textbf{S$^2$-MAD.} S$^2$-MAD proposes two approaches to identify agents with similar or identical viewpoints, which guides communication pruning decisions: (1) answer equivalence for tasks with a closed-form answer (e.g., math), and (2) response embedding similarity when answers are not directly comparable for equivalence (e.g., coding). We use (1) for both IMO-Answerbench (math) and HLE (multiple-choice) benchmarks.

\noindent\textbf{SID-ET.} SID-ET applies a threshold over a single agent's pre-debate min. LL, to determine if the question is solvable by a single agent, or requires all-to-all MAD. We sweep threshold values that lead to different skip rates (i.e., X\% questions are skipped from MAD due to sufficient single-agent min. LL, for X=\big\{10, 20, ..., 90\big\}). A higher skip rate reduces communication costs but may also reduce accuracy by aggressively skipping debates.

\begin{algorithm}
\caption{\label{alg:sid}Skip rate tuning for SID-ET.}
\begin{algorithmic}[1]
\Statex \textbf{Input:} (1) $Tok_{ref}$: Total token used by \sysname, (2) $Acc_{ref}$: Accuracy of \sysname.
\Statex \textbf{Output:} Selected skip rate $\hat{r}$

\State \textbf{Initialize} skip rate $r = 90\%$

\While{$r \geq 10\%$}
    \State $Tok_{SID}, Acc_{SID} = SID-ET(r)$
    \If{$Acc_{SID} \geq Acc_{ref}$}
        \State \Return $\hat{r} = r$
    \ElsIf{$Tok_{SID} > Tok_{ref}$}
        \State \Return $\hat{r} = r$
    \EndIf
    \State $r \mathrel{-}= 10\%$
\EndWhile

\State \Return $\hat{r} = 10\%$
\end{algorithmic}
\end{algorithm}

We tune SID-ET's skip rate per <LLM, dataset> setting. We aim to find an optimal skip rate that Pareto-dominates \sysname in terms of tokens and accuracy. If not found, we report results using the SID-ET setting that just exceeds the token cost of \sysname. This tuning procedure is documented in Algorithm~\ref{alg:sid}. Overall, no settings of SID-ET Pareto-dominates \sysname in terms of tokens and accuracy. The final selected skip rates are in 60/70 for IMO-AnswerBench/HLE using GPT-OSS-120B, and 50/60 using DeepSeek-V3.1.

Note that the original SID paper~\citep{Chen2025Sid} has a second technique, which compresses message size in all-to-all MAD based on token-level attention scores. This compression technique is triggered when early-stopping fails, and it only reduces token cost but not the number of communications. Our experiments use larger LLMs than \cite{Chen2025Sid} (i.e., hundreds vs. tens billions of parameters), making this attention-driven compression computationally expensive to evaluate. However, \sysname still saves in the number of cross-agent communications.

\noindent\textbf{\sysname.} We detail the configuration scheme for key hyperparameters in \sysname.
\begin{packeditemize}
\item Acceptance threshold ($\tau$): In our implementation, we accept an answer if the agent achieves 100\% SVR for at least $C$ challengers. We set $C=2$ for GPT-OSS-120B and $C=3$ for DeepSeek-V3.1. This is because agents in DeepSeek-V3.1 typically have more disagreements before debate: as shown by Table~\ref{tab:dataset-size}, for each dataset, using DeepSeek-V3.1 yields more non-unanimous pre-debate questions. Therefore, we set a higher acceptance threshold for DeepSeek-V3.1 agents to ensure disagreements are sufficiently evaluated.
\item Communications per round ($S$): We set $S=2$ for GPT-OSS-120B, and $S=3$ for DeepSeek-V3.1, based on their respective acceptance thresholds.
\item Communication budget per problem ($B_{max}$): We set $B_{\max} = S \cdot (k + m)$, where $k$ is the number of distinct pre-debate answer clusters and $m$ is the size of the largest cluster. The first term provides budget for one probed representative per cluster; the second adds budget to cover all members of the largest cluster. This simple rule works well in our experiments; a more principled budget allocation that further tightens costs is an interesting direction for future work.
\end{packeditemize}

\noindent\textbf{Additional implementation details.} Additional implementation decisions such as prompts used for initial response generation and MAD facilitation are in provided code implementations.

\section{\label{sec:ablation}Ablation Studies}

\begin{figure}
\centering
\includegraphics[width=\columnwidth]{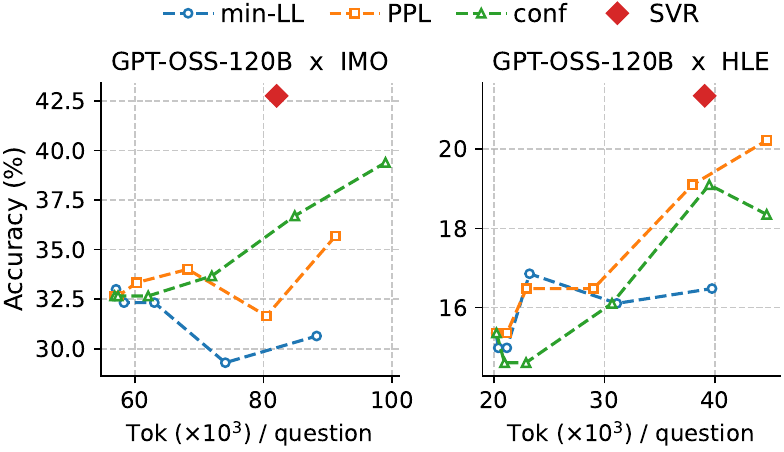}
\caption{\label{fig:ablation}\textbf{Ablation shows that SVR improves posterior scoring.} Token--accuracy tradeoffs when replacing SVR with alternative post-debate posterior signals, evaluated across acceptance rates from 10\% to 90\%.}
\end{figure}

We isolate the contribution of SVR by replacing it in Algorithm~\ref{alg:sysname} while keeping the same communication-control procedure. Specifically, we modify the posterior score computation on line~9. For agent $\mathcal{A}_r$, let $D$ denote the number of debates involving $\mathcal{A}_r$ as receiver, and let $\mathcal{P}$ denote the corresponding set of peer agents. We compute:
\[
\text{Posterior}(\mathcal{A}_r) =
\frac{1}{D}
\sum_{\mathcal{A}_s \in \mathcal{P}}
\text{Signal}\!\left(\text{Debate}(\mathcal{A}_s, \mathcal{A}_r)\right).
\]

Here, $\text{Signal}$ is one of min. LL, PPL, or self-reported confidence, extracted from $\mathcal{A}_r$'s post-debate responses. Thus, instead of using SVR to measure whether $\mathcal{A}_r$ retains its answer after debate, these ablations use post-debate confidence-like signals averaged across all debates.

Figure~\ref{fig:ablation} shows the token--accuracy tradeoff on IMO-AnswerBench and HLE using GPT-OSS-120B. We compare the default \sysname, which uses SVR as the posterior score, against variants that replace SVR with alternative post-debate signals. For each variant, we sweep the acceptance threshold to cover acceptance rates from 10\% to 90\%.

Overall, SVR demonstrates more optimal token--accuracy tradeoffs than alternative posterior score choices. SVR achieves higher accuracy than alternatives that fail to match it under all tested acceptance rates, with +3.37 and +1.13 points on IMO and on HLE compared to the strongest competitor. At matched token cost, SVR improves accuracy by +8.75 and +2.25 points compared to the best alternative signal on the two datasets, respectively. This shows that SVR is an effective posterior signal independent of the communication control-flow in Algorithm~\ref{alg:sysname}.

\section{Use of AI Assistants}

We used AI assistants for writing polish and code development support. All research ideas, methodological decisions, experimental design, analysis, and conclusions are the authors' own work.

\end{document}